\documentclass[reprint,showpacs,aps,prb,amsmath,amssymb]{revtex4-1}
\date{\today}
\usepackage{amssymb}
\usepackage{amsmath}
\usepackage{mathrsfs}
\usepackage{graphicx}
\usepackage{dcolumn}
\usepackage{color}

\begin{document}

\title{Fractional exclusion statistics and thermodynamics of the Hubbard chain in the spin-incoherent Luttinger liquid regime}
\author{Carlindo Vitoriano$^1$, R. R. Montenegro-Filho$^2$, and M. D. Coutinho-Filho$^2$}
\affiliation{$^1$Unidade Acad\^emica do Cabo de Santo Agostinho, Universidade
Federal Rural de Pernambuco, 54503-900 Cabo de Santo Agostinho-PE, Brazil\\
$^2$Laborat\'orio de F\'{\i}sica Te\'orica e
Computacional, Departamento de F\'{\i}sica, Universidade Federal
de Pernambuco, 50670-901 Recife-PE, Brazil}

\begin{abstract}
Bethe ansatz and bosonization procedures are used to describe the thermodynamics of the strong-coupled Hubbard chain in the 
\textit{spin-incoherent} Luttinger liquid (LL) regime: $J(\equiv 4t^2/U)\ll k_B T\ll E_F$, where $t$ is the hopping amplitude,
$U(\gg t)$ is the repulsive on-site Coulomb interaction, and $k_B T (E_F\sim t)$ is the thermal (Fermi) energy.
We introduce a fractional Landau LL approach, whose $U=\infty$ fixed point is exactly mapped 
onto an ideal gas with two species obeying the Haldane-Wu \textit{exclusion} fractional statistics.
This phenomenological approach sheds light on the behavior of several thermodynamic
properties in the spin-incoherent LL regime: specific heat, charge compressibility,
magnetic susceptibility, and Drude weight. In fact, besides the hopping (mass) renormalization, 
the fractional Landau LL parameters, due to quasiparticle interaction, are determined and relationships with velocities of
holons and spinons are unveiled. The specific heat thus obtained is in very good agreement with 
previous density matrix renormalization group (DMRG) simulations of the $t$-$J$ model in the spin-incoherent regime. 
A phase diagram is provided and two thermodynamic paths to access this regime clarifies both 
the numerical and analytical procedures.
Further, we show that the high-$T$ limit of the fractional Landau LL entropy and chemical potential 
exhibit the expected results of the $t$-$J$ model, under the condition $U\gg k_B T$.
Lastly, finite-temperature Lanczos simulations of the single-particle distribution function confirm the characteristics of the 
spin-incoherent regime and the high-$T$ limit observed in previous DMRG studies.
 \end{abstract}

\maketitle

\section{Introduction}
Very recently, experimental realization of one-dimensional (1D) ultracold fermions with tunable number of 
spin components has been reported in the crossover regime of temperature between spin-ordered and spin-incoherent 
Luttinger liquid (LL) \cite{Pagano2014}. In particular, the
subtle bosonic limit \cite{Yang11} is evidenced for strongly repulsive $^{173}$Yb atoms with nuclear spin $I=5/2$. In addition, studies using analytical 
and numerical methods have shown \cite{Jen2016} that the spin-incoherent 1D spin-1 Bose
LL in a harmonic trap and in the Tonks-Girardeau limit (infinite repulsion) \cite{Tonks1936,*Girardeau1960,*Lieb1963}, 
exhibits the universal $1/p^4$ dependence momentum distribution, which is, however, 
broader than the spinless case, due to spin-function overlaps. 
We also remark that the Tonks-Girardeau limit has been experimentally achieved in ultra cold boson atoms \cite{Kinoshita1125,*Paredes2004,*Haller1224}, and also
verified in frustrated quantum spin chains \cite{Montenegro-Filho2008}. 

On the theoretical side, the method of bosonization \cite{stone} has provided an efficient means to derive analytical results for low-dimensional
interacting fermion systems in condensed matter and field theory, thereby allowing the emergence of new physical concepts. In this context, the
LL theory has been proposed \cite {142585} as a unified framework to describe the low-energy physics of a large class of
1D quantum many-body systems \cite{giamarchi,essler2005one, takahashi}. Emphasis has been given to those systems subjected to strong
quantum fluctuations and exhibiting new features not fully described by the standard Fermi liquid theory \cite{pines} governed by the
zero coupling-strength fixed point \cite{66129}. Notwithstanding, several aspects of a Landau-Luttinger theory were discussed at length
\cite{3757,449967,73b}. Further, generalization of the standard Fermi liquid theory was also proposed with aim 
in describing the unusual properties of heavy-fermion systems, in particular close to a metal-insulator transition \cite{Shaginyan2010}.

Following the LL concept we have witnessed a vigorous development in the study of 1D strongly correlated electron systems, particularly in
connection with the nature and the role played by charge and spin excitations, and the related phenomenon of spin-charge separation
\cite{giamarchi}. Comparison of results derived using bosonization with those from other methods, such as the Bethe-ansatz and 
density matrix renormalization group techniques 
\cite{takahashi,Sacramento2013}, has also proved valuable. More recently, a very interesting regime of the LL, namely the spin-incoherent LL, has received special attention
\cite{79801}. For both continuous \cite{92b,93226401} and lattice \cite{394845,81075108,106146401} versions of the 1D Hubbard model \cite{essler2005one}, this regime
is realized under the condition $J(\equiv 4t^2/U)\ll k_B T\ll E_F(\sim t)$, where $t$ is the nearest-neighbor hopping amplitude,
$U$ is the repulsive on-site Coulomb interaction, $\beta=1/(k_BT)$ is the inverse temperature measured in units of the Boltzmann constant, $J$ is the antiferromagnetic exchange
coupling, and $E_F$ is the Fermi energy. Alternatively, for low carrier densities, quantum wires \cite{92a,72045315,fieteprb2005,101036801,22095301} are near the 1D Wigner
crystal limit at which the electrostatic energy between the particles greatly exceeds their kinetic energy leading to $J\ll E_F$, so that for $k_BT\gg J$ the observed conductance is
about half the usual LL value $2e^2/h$ due to the spin-incoherent contribution to the resistance, where $e$ is the magnitude of electron charge and $h$ is the Planck constant. 
Indeed, it has been shown that, despite features of spin-charge separation persist, the spin part of the correlation function exhibits an exponential
spatial decay \cite{92b,93226401}
not consistent with the usual LL power-law decay. Moreover, at half filling \cite{394845}, the effective gapped charged excitations are modified due
to the presence of the uncorrelated spin degrees of freedom. 

In this work we shall demonstrate that the thermodynamic properties of the Hubbard chain in the spin-incoherent regime
can be described by using arguments from complementary powerful methods in the realm of quantum statistical mechanics and 
many-body theory, notably the Haldane-Wu {\it exclusion} fractional statistics \cite{HWF}. 
In this context, the fractional character of the excitations of Hubbard models
with short-range Coulomb interaction and correlated hopping \cite{102146404,allegra2011,hidden2016} (bond-charge interaction), 
and infinite-range Coulomb interaction \cite{PhysRevB.61.7941,*PhysRevB.72.165109} as well, 
has been invoked to properly describe phase diagrams exhibiting metal-insulator transition, including the unexpected absence of conductivity 
at half filling due to a topological 
change in the Fermi surface, and $\eta$-pairing \cite{PhysRevLett.63.2144} induced 1D critical superconductivity \cite{82125126}.
Correlated hopping can also play a relevant role in 2D models of high-temperature superconductors \cite{universal2017}.
In addition, particles obeying exclusion fractional statistics have been considered in the context of optical lattices \cite{keilmann2011,optics2012}, 
including the (1D) Tonks-Girardeau limit  \cite{Batchelor2006}.
In 2D systems, it was suggested \cite{Cooper} that spectroscopy measurements on ultracold atoms can be used 
to demostrate the fractional exclusion statistics of quasiholes in the Laughlin state of bosons.
On the other hand, neutral anyonic excitations, which satisfy fractional
\textit{exchange} statistics in two dimensions, can be identified \cite{CES} through measurements of spectral functions near the threshold.
The structure factor follows a universal power-law behavior, whose exponent is the signature of the anyon statistics and the 
underlying topologically ordered states that should occur in spin liquids and fractional Chern insulators.
Moreover, it was proposed \cite{Arcila-Forero2018} that superfluid to Mott insulator quantum phase transitions 
in an anyon-Hubbard model with three-body interaction can be driven by the statistics or by the interaction. 

In Sec. \ref{sec:strong-coupling}, we use a strong-coupling perturbative expansion
\cite{46a,*ha1996quantum} of the Takahashi's Bethe-ansatz grand-canonical free energy \cite{4769,52103,65165104} to calculate the 
Helmholtz free energy, energy and entropy in the spin-incoherent regime.
From these thermodynamic potentials and the Luttinger theory, we present in Sec. \ref{sec:incohell} the
specific heat, isothermal compressibility, Luttinger liquid parameter, magnetic susceptibility, and the Drude weight,
to leading order in $J/E_{F}$. 
In Sec. \ref{sec:uinfity}, we show that the thermodynamics
of the infinite-$U$ Hubbard chain is exactly mapped onto an ideal excluson gas of two species obeying the Haldane-Wu
{\it exclusion} fractional statistics \cite{HWF}. 
In Sec. \ref{sec:flll} we introduce a fractional Landau LL approach, which provides non-trivial insights and a direct connection with 
the LL theory in the spin-incoherent regime. Indeed, our results provide strong evidence that the 
fractional excluson entropy describes very well the thermodynamics of the spin-incoherent regime.
We can thus identify the pertinent fractional Landau LL parameters, and their 
relationship with the LL properties, namely, the velocity of holons and spinons. 
Despite that there have been previous attempts \cite{242130,Shaginyan2010,Leinaas2017} towards a generalization of
the Fermi liquid theory to particles obeying fractional exclusion statistics, a realization of these ideas, as presented here, is
apparently missing. In Sec. \ref{sec:hight} we consider the high-$T$ limit \cite{Juttner1997} of the particle distribution function, 
chemical potential and entropy. Finally, concluding remarks are reserved to Sec. \ref{sec:remarks}.

\section{Spin-Incoherent Regime of the Hubbard chain}
\label{sec:strong-coupling}
The Hamiltonian of the Hubbard chain of $L$ sites in the presence of an external magnetic field along
the $z$ direction is given by
\begin{eqnarray}
{\cal H}=-t\sum_{\langle
i,j\rangle,\sigma}
c_{i\sigma}^{\dagger}c_{j\sigma}^{}
+U\sum_{i}n_{i\uparrow}n_{i\downarrow}
 -\mu_B H\sum_{i}(n_{i\uparrow} -
n_{i\downarrow}),
\nonumber\\
\label{modelo}
\end{eqnarray}
where $\langle i,j\rangle$ denotes nearest-neighbor sites, $\sigma\in\{\uparrow, \downarrow\}$,
$c^{}_{i\sigma}~(c_{i\sigma}^{\dagger})$ is the electron annihilation (creation) operator, $n_{i\sigma}^{}=
c_{i\sigma}^{\dagger}c_{i\sigma}^{}$ is the number operator, $\mu_BH$ is the Zeeman energy, and $\mu_B$ is the Bohr magneton.
The $t-J$ model, which projects out doubly occupied states in the strong-coupling regime of the Hubbard chain, reads:
\begin{eqnarray}
 \mathcal{H}_{t-J}&=&-t\sum_{\langle i,j\rangle,\sigma}(1- n_{i\bar{\sigma}}) c_{i\sigma}^{\dagger}c_{j\sigma}^{}(1- n_{j\bar{\sigma}})\nonumber\\
& &+J\sum_{\langle i,j\rangle}\left({\mathbf{S}}_{i}\cdot {\mathbf{S}}_{j}-\frac{1}{4}n_in_j\right)-2\mu_B H S^z,
\label{eq:tj}
\end{eqnarray}
where $\bar{\sigma}=-\sigma$, $S^z=\frac{1}{2}\sum_{i}(n_{i\uparrow} - n_{i\downarrow})$, with $\hbar\equiv1$, and
$$J=4t^2/U.$$

The spin-incoherent LL regime is found at temperatures such that
\begin{equation}
J(\equiv 4t^2/U)\ll k_{B}T \ll E_{F}\sim t.
\label{siregime}
\end{equation}
This regime is characterized by low-energy collective charge
excitations (holons) with a velocity $v_c^{(inch)}$ of interacting spinless 
fermions, and by
the absence of collective spin excitations, since the very small strong-coupling spinon velocity $v_s$ ($\sim J$) implies a 
very small correlation length $\xi=v_s/\pi k_B T\sim J/2k_BT \ll 1$. In this context, we note that the special point $J=0$ ($U=\infty$) is 
also a spin-incoherent LL, since it is a spin-disordered state, with $v_s=J=0$ and infinite spin degeneracy in the thermodynamic limit; 
thereby, only holon excitations exist.

The thermodynamic Bethe ansatz has been successfully implemented for the Hubbard chain long ago
\cite{4769}. However, difficulties exist in deriving closed-form expressions for thermodynamic quantities from the
 infinite coupled integral equations. Notwithstanding, it has been shown \cite{46a} that it is possible to solve the set
of integral equations perturbatively in the strong coupling limit ($t\ll U$), and consistent high-temperature 
series expansions have been provided. In particular, in Appendix \ref{strHa} the results reported 
in Ref. \onlinecite{46a} for the grand canonical free energy $\Omega(T,\mu,H)$ can be used in order to obtain corrections of ${\cal O}(t^2/U)$ to the $U=\infty$ limit.
Most importantly, as we show in this work, these corrections are suitable to describe the $t-J$ limit of the Hubbard chain in the 
regime $U\gg k_BT$, including the spin-incoherent regime for $k_BT\ll t$.
In fact, in Appendix \ref{strHa} we find that $\Omega(T,\mu,H)$ in the spin-incoherent regime reads:

\begin{eqnarray}
&&\frac{\Omega_{inch}(T,\mu,H)}{L}=\nonumber\\
&&-k_BT\int^{\pi}_{-\pi}\frac{dk}{2\pi}\ln[1+e^{-\beta(\varepsilon_k-\mu-\mu_BH)}
+e^{-\beta(\varepsilon_k-\mu+\mu_BH)}]
\nonumber\\
&&-\frac{k_BT}{\cosh(\beta \mu_BH)}\left(\frac{t}{U}\right)
\int^{\pi}_{-\pi}\frac{dk}{2\pi}\frac{2}{e^{\beta(\varepsilon_k-\mu)}+2\cosh(\beta \mu_BH)}
\nonumber\\
 &&\times \int^{\pi}_{-\pi}\frac{dk}{2\pi}\cos{k}\ln[1+e^{-\beta(\varepsilon_k-\mu-\mu_BH)}
+e^{-\beta(\varepsilon_k-\mu+\mu_BH)}]
\nonumber\\
&&+\cdots,
\label{tanto}
\end{eqnarray}
where $\mu$ is the chemical potential and $\varepsilon_k=-2t\cos{k}$ is the 
dispersion relation of tight-binding fermionic particles, which is the exact dispersion 
relation for the $U=\infty$ case\cite{1571033}. In fact, making $U=\infty$ in 
Eq. ($\ref{tanto}$), we obtain the exact expression of the grand-canonical free energy \cite{4769} 
at this extremal coupling value. The grand-canonical free energy (\ref{tanto}) is also suitable to describe the spin-incoherent 
regime, since using the inequalities in (\ref{siregime}): $4t^2/U \ll k_BT\ll t$, we find $U/k_BT\gg 1/(k_BT/t)^2\gg 1$.

The chemical potential $\mu$ is calculated from
$n=-\frac{1}{L}\left(\frac{\partial\Omega}{\partial\mu}\right)$:
\begin{eqnarray}
&&\mu_{inch}(T,n)= -2t\cos(n\pi)
 \nonumber\\
&&-\frac{nt^2}{U}\left[1+ 2\sin^2(n\pi)
 -\frac{\sin(2n\pi)}{2n\pi}\right]
 -k_BT\ln{2}
\nonumber\\
&&+\frac{\pi^2(k_BT)^2\cos(n\pi)}{12t\sin^2(n\pi)}
 \left\{1+\left(\frac{2t}{U}\right)
\left[\frac{n}{\cos(n\pi)}-\frac{\sin(n\pi)}{\pi}
\right]\right\}
\nonumber\\
&&+\cdots,
\label{muinch}
\end{eqnarray}
The corresponding expansion for the Helmholtz free energy $F(=\mu N+\Omega)$, energy $E(=F-T\partial F/\partial T)$, and 
entropy $S(=-\partial F/\partial T)$ read:
\begin{eqnarray}
&&\frac{F_{inch}(T,n)}{L}=-\frac{2t\sin(n\pi)}{\pi}
 -\left( \frac{t^2}{U}\right)n^2\left[1-\frac{\sin(2n\pi)}{2n\pi}\right]
 \nonumber\\
&&-nk_BT\ln{2}-\frac{\pi(k_BT)^2}{12t^\ast \sin(n\pi)}
+\cdots;
\nonumber\\
 \label{pasta}
\end{eqnarray}
 
 \begin{eqnarray}
\frac{E_{inch}(T,n)}{L}&=&-\frac{2t\sin(n\pi)}{\pi}
 -\left( \frac{t^2}{U}\right)n^2\left[1-\frac{\sin(2n\pi)}{2n\pi}\right]
 \nonumber\\
 &+&\frac{\pi(k_BT)^2}{12t^\ast \sin(n\pi)}
+\cdots;
\label{pastel}
\end{eqnarray}
 \begin{equation}
\frac{S_{inch}(T,n)}{L}= nk_B\ln{2}+\frac{\pi k_B^2 T } {6t^\ast\sin(n\pi)}+\cdots,
\label{next}
\end{equation}
where the $T$-dependent terms have coefficients with a 
hopping parameter $t^{\ast}$ given by, up to ${\cal O}(t/U)$, 
\begin{equation}
t^{\ast} = t\left[1-\frac{2nt\cos(n\pi)}{U}\right].
\label{za}
\end{equation}
We stress that up to ${\cal O}(t/U)$ doubly occupied sites are forbidden \cite{377541}. In fact,
 \begin{eqnarray}
 \frac{\langle N_{\uparrow\downarrow}\rangle}{L}&=&\frac{\partial(E_{inch}/L)}{\partial U}=
 n^2\left( \frac{t}{U}\right)^2\left[1-\frac{\sin(2n\pi)}{2n\pi}\right]
 \nonumber\\
 &-&\left( \frac{n\pi}{6}\right)\left(\frac{k_BT}{U} \right)^2\cot(n\pi)+\cdots.
 \label{papel}
 \end{eqnarray}
 
The above results show that the charge degrees of freedom in the regime $J\ll k_{B}T \ll t$ 
or $J=0$ and $k_BT\ll t$ are described by a gas of free spinless fermions.
Indeed, the first term in $E_{inch}(T,n)$ is the ground-state energy
of a gas of free spinless fermions
with dispersion $\varepsilon_k=-2t\cos{k}$;
while $T$-dependent terms in $E_{inch}(T,n)$ and $S(T,n)$ are contributions from thermally excited spinless
fermions, with a mass $\sim 1/t^\ast$, above the Fermi 
surface, which is defined by the wave vectors $k=\pm k_{F}$, with $k_{F}=n\pi$.
The spin-incoherent regime is identified by noticing that the first 
term in the entropy $S_{inch}(T,n)$ indicates that the spin degrees of freedom 
are fully disordered.

\section{Response Functions and Spin-incoherent LL parameters}
\label{sec:incohell}
The Hamiltonian of the system in the spin-incoherent regime and \textit{zero field} 
can be mapped onto the following
charged bosonized LL Hamiltonian \cite{fieteprb2005}:
 \begin{equation}
 {\cal H}_{\text{inch}}=v^{(inch)}_c\int \frac{dx}{2\pi}\left[\frac{1}{g}(\partial_x\theta)^2+g(\partial_x\phi)^2 \right],
 \label{incoh}
 \end{equation}
 where $v_c^{(inch)}$ is the holon velocity, $(1/\pi)(\partial_x\theta)$ is the
 fluctuation in electron density and the commutation relation $[\theta(x),\partial_{x'}\phi(x')]=i\pi\delta(x-x')$ holds.
 The coupling $g$ can be written in terms of the 
LL parameter $K_c$, which governs the decay of the correlation functions: 
\begin{equation}
K_c=\frac{1}{2g}.
\label{eq:kc}
\end{equation}

The specific heat $C=-\frac{T}{L}\left(\partial^2F/\partial T^2\right)$:
 \begin{eqnarray}
C_{inch}(T,n)=\gamma_{inch}k_B^2 T+\cdots,
\label{sexta}
\end{eqnarray}
displays a free spinless Fermi gas form where the specific-heat coefficient $\gamma_{inch}$ and the holon velocity 
are, respectively,
\begin{equation}
\gamma_{inch}=\frac{\pi}{3v^{(inch)}_{c}};
\label{ginch}
\end{equation}

\begin{equation}
v^{(inch)}_c =2t^\ast\sin(n\pi).
\label{girl}
\end{equation}

On the other hand, the charge compressibility
$\kappa^{-1}=n^2(\partial \mu / \partial n)$ reads:
\begin{eqnarray}
&&\kappa_{inch}^{-1}(T,n)=2\pi t n^2 \sin(n\pi)
\nonumber\\
&&\times \left\{ 1-\left( \frac{2t}{U}\right)\left[ \frac{\sin(n\pi)}{\pi}+n\cos(n\pi)+{\cal O}\left(\frac{k_B^2 T^2}{t^2}\right)\right]\right\}\nonumber\\
&&
\label{indio}
\end{eqnarray}
Further, in the spin-incoherent LL regime $g_{inch}^{-1}=\pi v^{(inch)}_{c}\kappa_{inch} n^{2}$, we find
\begin{equation}
g_{inch}=1-\left(\frac{2t}{U}\right) \frac{\sin(n\pi)}{\pi},
\end{equation}
and
\begin{equation}
 K^{(inch)}_{c}=\frac{1}{2g_{inch}}=\frac{1}{2}+\left(\frac{t}{U}\right)\frac{\sin(n\pi)}{\pi}.
 \label{eq:kci}
\end{equation}
Notice that using Eqs. (\ref{za}) and (\ref{girl}), we can verify that 
$v_c^{(inch)}$ is not the holon velocity of the standard LL theory at $T=0$.

Lastly, since \cite{schulz} $\sigma_0=2K_c v_c$, the Drude weight that measures the dc peak in the conductivity,
$\sigma(\omega)=\sigma_0\delta(\omega)$,
 in the spin-incoherent LL regime is given by
 \begin{equation}
 \sigma_0^{(inch)}=2t\sin(n\pi)\left[
 1+\frac{2t}{U}\left(\frac{\sin(n\pi)}{\pi}-n\cos(n\pi)\right)\right],
 \label{trans}
 \end{equation}
where use was made of Eqs. (\ref{eq:kci}) and (\ref{sv}).

We also confirm the spin-incoherent regime by probing the spin degrees of freedom through the susceptibility 
$\chi(T,\mu)$.  As shown in Appendix \ref{susInch}, the canonical susceptibility
and spinon velocity read, respectively:

\begin{eqnarray}
\chi_{inch}(T,n)= { \mu_B^2\beta n} \left[1-\frac{n v_s}{\pi k_BT} +{\cal  O}\left(\frac{J}{t}\right)\right];
\label{graviola}
\end{eqnarray}

 \begin{equation}
 v_s=\frac{2\pi t^2}{U}\left[1-\frac{\sin(2n\pi)}{2n\pi}\right],
 \label{sv}
 \end{equation}
 where $v_s$ is the strong-coupling spinon velocity \cite{schulz}.
The correction of ${\cal O}(v_s/k_BT)$ to the
dominant Curie response is the one we expect in view of the highly excited
spin degrees of freedom, and implies $v_s(n)|_{U=\infty}=0$, for any value of $T$.
For finite $J$, we use the fluctuation-dissipation theorem: $\chi=\beta\int G(x)\,dx$, 
where $G(x)$ is the spin-correlation function. In order to satisfy Eq. (\ref{graviola}),
$G(x)=\mu_B^2n[\delta(x)-ne^{-x/\xi}]$, with a correlation length $\xi$ 
given by the expected result \cite{142585,Hirokazu1991,Klumper1996}: 
$\xi=v_s/(\pi k_BT)\sim [J/(2k_BT)] \ll 1$, thus confirming 
the spin-incoherent regime for finite $J$. 

\subsection{$T\rightarrow 0$ limit: the standard LL regime, with charge and spin collective excitations}
Here we show that we can infer the parameters of the standard LL regime, which 
settles as $T\rightarrow0$, from the above spin-incoherent results.
In doing so, we take advantage of the description of the $U\rightarrow\infty$ limit of
the Hubbard chain put forward in Ref. \onlinecite{5515475}. In particular, 
by using the Bethe ansatz solution, it has been shown that the ground-state wave function of the
system can be constructed as a product of a spinless fermion wave function $|\Psi\rangle$ and a squeezed spin wave function $|\chi\rangle$.
The wave function $|\chi\rangle$ are eigenfunctions of the following Heisenberg Hamiltonian:
 \begin{equation}
 \label{Heisenbergnew}
 {\cal H}_{S}=\sum_{i=1}^{N}\sum_{\alpha=x,y,z}\tilde{J}^{\alpha}
 \left(S_{i}^{\alpha}S_{i+1}^{\alpha}-\frac{1}{4}\delta_{\alpha,z}\right),
  \end{equation}
where
\begin{equation}
\label{Jnew}
 \tilde{J}^{\alpha}=n\frac{4t^{2}}{U}\left[1-\frac{\sin\left(2n\pi\right)}{2n\pi}\right]
\end{equation}
is determined by the ground-state energy wave function of the spinless fermions $|\Psi^{GS}\rangle$. Notice that, at
half filling, we have the standard coupling $J=4t^{2}/U$.
Therefore, the contribution of $\cal H_{S}$ to the ground-state
energy per site is given by
\begin{eqnarray}
\label{GSenergy}
 \frac{\langle \chi^{GS}|{\cal H}_{S}| \chi^{GS} \rangle}{L}&\equiv&\frac{E^{GS}}{L}=-n^{2}
 \left(\frac{4t^{2}}{U}\right)\frac{\left[1-4\gamma_{S}\left(T=0\right)\right]}{4}
 \nonumber\\
 &\times&\left[1-\frac{\sin\left(2n\pi\right)}{2n\pi}
 \right],
\end{eqnarray}
where
\begin{eqnarray}
\gamma_S(T)=\langle {\bf S}_i\cdot {\bf S}_{i+1}\rangle
=\left\{
\begin{array}{r}
 1/4-\ln{2}, \quad T=0 ;\\
 0, \quad k_BT\gg t^2/U,
 \end{array}
 \right.\label{gammaS}
 \end{eqnarray}
 denotes the $T$-dependent nearest-neighbor spin correlation function of the Heisenberg
 model \cite{38363}.
 This contribution at $T=0$, together with that of spinless
 fermions [first term in Eq.~(\ref{pastel})] is the exact
 ground-state result up to ${\cal O}(t/U)$ \cite{6930,412326,641831,5515475,2351196,*anderson2017theory} of the 
 1D $t$-$J$ model. We thus
 infer that the ground state energy of the Hubbard chain in the
 spin-incoherent regime obtains through the replacement of $\gamma_{S}\left(T=0\right)$
 by $\gamma_{S}\left(T\gg J/k_{B}\right)=0$. This correspondence was already
 noticed in the study of the thermodynamics of the Hubbard chain in the
 spin-disordered regime at half filling \cite{394845}.
 

We have also noted
that several expressions valid in the spin-incoherent LL regime differ
from the corresponding ones at $T=0$ by the multiplying factor $[1-4\gamma_S(T=0)]$.
\begin{figure}
\begin{center}
\includegraphics*[width=0.47\textwidth]{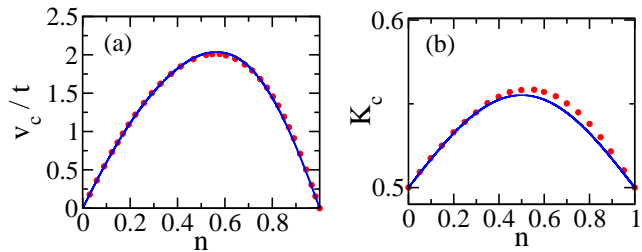}
\caption{(color online). (a) Charge velocity [Eq. (\ref{girls})] and (b)
correlation exponent $K_c$ [Eq. (\ref{noises})] at $T=0$ as a function of $n$ for $U=16t$. In both figures,
the dots displayed were obtained from Ref.~\onlinecite{schulz}.} \label{patrol}
\end{center}
\end{figure}

Consider first the charge velocity at $T=0$:
\begin{eqnarray}
v_c(T=0,n)&=&2t\sin(n\pi)
\nonumber\\
&&\times \left \{1-\frac{2[1-4\gamma_S(0)]nt\cos(n\pi)}{U}\right\},\nonumber\\
&=&2t\sin(n\pi)\left [1-\frac{8\ln{2}}{U}nt\cos(n\pi)\right]
\label{girls}
\end{eqnarray}
which is the extension of Eq.~(\ref{girl}) to $T=0$ using Eq. (\ref{gammaS}),
in agreement with Bethe-ansatz analytical results \cite{*[][{. Notice a factor of 2 discrepancy, in the $(t/U)$ correction, for the prediction of $v_c$: our Eq. (\ref{girls}) and Eq. (6.37) 
of this citation.}] pencsol} of the 
strongly coupled Hubbard model at $T=0$.
In Fig.~\ref{patrol}(a) we plot $v_c(T=0)$ as a function of $n$ for $U=16t$. 
Note the remarkable agreement with early Bethe-ansatz numerical \cite{schulz} result at $T=0$.

Now, consider the LL parameter at $T=0$:
 \begin{eqnarray}
K_{c}(T=0,n)&=&\frac{1}{2}+ [1-4\gamma_S(0)]\left(\frac{t}{U}\right) \frac{\sin(n\pi)}{\pi}\nonumber\\
&=&\frac{1}{2}+ \frac{4\ln{2}}{U\pi}t\sin(n\pi).
\label{noises}
\end{eqnarray}
The validity of this formula is confirmed in Fig.~\ref{patrol}(b), where the plot of
$K_{c}(T=0,n)$ as a function of $n$ for $U=16t$ is exhibited. 
In addition, we note that for $n\rightarrow 0$:
$K_{c}(T=0,n)=1/2+(4\ln{2})(nt/U)$, which coincides with the expression for $K_c$ reported
in Ref.~\onlinecite{79195114}.
 
\begin{figure}
\begin{center}
\includegraphics*[width=0.27\textwidth]{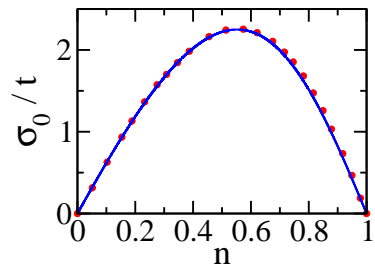}
\caption{(color online). Drude weight as a function of bandfilling for $U=16t$ and
$T=0$. Solid curve is the plot of Eq.~(\ref{dru}) and the dots in highlight
were obtained from Ref.~\onlinecite{schulz}. } \label{gato}
\end{center}
\end{figure}

The previous results imply that the Drude weight \cite{*[{For $T^2$ contribution, see: }] [] Fujimoto1998} 
at $T=0$ is given by
\begin{eqnarray}
\sigma_0(T=0)=2K_c v_c=2t\sin(n\pi)\left\{1+8\ln{2}\left(\frac{t}{U}\right)
\right. \nonumber\\
\left. \times\left[\frac{\sin(n\pi)}{\pi}-n\cos(n\pi)\right]\right\},
\label{dru}
\end{eqnarray}
where use of Eqs.~(\ref{girls}) and (\ref{noises}) has been made.
As shown in Fig.~\ref{gato}, the agreement between this formula
for $U=16t$ and early numerical results \cite{schulz} is excellent.

Lastly, concerning the specific-heat coefficient, as $T\rightarrow0$
 the spin-spin correlation function displays power-law behavior
 and the prediction for $\gamma$ is \cite{essler2005one}:
\begin{equation}
\gamma =\frac{\pi}{3}\left(\frac{1}{v_c}+\frac{1}{v_s}\right)_{T=0}.
\label{gammat0}
\end{equation}

\section{$U=\infty$ as an exact ideal gas of exclusons or free spinless fermions}
\label{sec:uinfity}
The concept of a Luttinger liquid is the paradigm for describing the low-energy physics of interacting electron systems
in one dimension. Notwithstanding, it is important to investigate alternative approaches that can shed light
on the physics of such systems. In this context, a remarkable result
that follows from previous works \cite{102146404,82125126,PhysRevB.61.7941,*PhysRevB.72.165109} by two of the authors is that the properties
of $U=\infty$ limit can be viewed as derived from
an ideal excluson gas of two fractional species: $\alpha=1$ for particles
with spin up and $\alpha=2$ for particles with spin down, coupled by the Haldane statistical matrix
\begin{eqnarray}
[g]_{{ k}{ k'};\alpha\alpha'}=\delta_{{ k}{ k'}}\left(
\begin{array}{cc} 1&1\\ 0&1\end{array}\right),
\label{canto}
\end{eqnarray}
in which case double occupation is excluded. 
In fact, the same $3\times 3$ statistical matrix describes the referred Hubbard
models \cite{102146404,82125126,PhysRevB.61.7941,*PhysRevB.72.165109}, including double occupancy effects.
This is confirmed by noting that Eq.~(\ref{tanto}) with $U=\infty$ can be written in the form:
\begin{equation}
\Omega_{\infty}(T,\mu_{\infty},H)=-\frac{1}{\beta}\sum_{k,\alpha}\ln(1+w_{k,\alpha}^{-1}),
\end{equation}
where $w_{k,\alpha}$'s satisfy the Haldane-Wu distribution \cite{HWF}:
\begin{eqnarray}
w_{k,1}&=&e^{\beta({\varepsilon}_{k,1}-\mu_{\infty})}, \\
w_{k,2}&=&(1+w_{k,1})e^{\beta({\varepsilon}_{k,2} -
{\varepsilon}_{k,1})}.
\end{eqnarray}
In addition, $\langle{n}_{{k},\alpha}\rangle$ satisfies the exclusion relation:
\begin{equation}
\langle{n}_{k,\alpha}\rangle
w_{k,\alpha}=1-\sum_{k',\lambda}g_{kk';\alpha\lambda} \langle
{n}_{k',\lambda}\rangle,
\end{equation}
 where
 \begin{eqnarray}
\langle{n}_{k,\alpha}\rangle=\frac{e^{-\beta({\varepsilon}_{k,\alpha}-\mu_{\infty})}}
{{\displaystyle 1+\sum_{\lambda=1}^{2}e^{-\beta({{\varepsilon}}_{k,\lambda}-\mu_{\infty})}}}.
\label{mesa}
\end{eqnarray}
More specifically:
\begin{eqnarray}
\langle n_{k,1}\rangle &=&e^{2\beta\mu_B H}\langle n_{k,2}\rangle, \label{eq:excinf1}\\
&=& \frac{e^{\beta\mu_B H}}{e^{\beta(\varepsilon_k-\mu_\infty)}+2\cosh(\beta\mu_B H)},\label{eq:excinf2}
\end{eqnarray}
in agreement with an independent calculation for the Hubbard model at $U=\infty$ in Ref. \onlinecite{pronko}. 
Although the matrix given in Eq.~(\ref{canto}) is asymmetric, it should be noted that the spin-up
and spin-down symmetry is preserved, as we can see from Eq.~(\ref{mesa}):
 $\langle{n}_{k,1} \rangle_{H}=\langle{n}_{k,2} \rangle_{-H}$.
Moreover, the entropy reads:
\begin{eqnarray}
&&S_{\infty}(T,\mu,H)=-k_B\sum_{k}[\langle{n}_{k,1}\rangle\ln{\langle{n}_{k,1}\rangle}
+\langle{n}_{k,2}\rangle\ln{\langle{n}_{k,2}\rangle}
\nonumber\\
 &&+(1-\langle{n}_{k,1}\rangle-\langle{n}_{k,2}\rangle)
\ln{(1-\langle{n}_{k,1}\rangle-\langle{n}_{k,2}\rangle)}],
\label{peso}
\end{eqnarray}
which carries the signature of the statistical matrix in Eq. (\ref{canto}).

In zero field, Eq.~(\ref{mesa}), or Eqs. (\ref{eq:excinf1}) and (\ref{eq:excinf2}) reduces to
\begin{equation}
\langle{n}_{k,1}\rangle_{H=0}=\langle{n}_{k,2}\rangle_{H=0}
=\frac{1}{e^{\beta({\varepsilon}_{k}-\mu_\infty)}+2}\equiv \langle n_{k}\rangle;
\label{money}
\end{equation}
in agreement with early results \cite{377541}, so $\langle n_{k}\rangle$ develops a
rigorous step discontinuity at the Fermi surface as $T\rightarrow 0$, with
\begin{equation}
n =\frac{2}{L}\sum_{k}\langle{n}_{k}\rangle_{T=0}.
\end{equation}
We also mention that the fractional character of $\langle n_{k}\rangle$, Eq.~(\ref{money}), stems
from the fact that, in the exclusion formalism, both charge and spin degrees of freedom are combined
to form a single distribution.
However, by summing up in the fractional species, we obtain the free spinless fermion distribution:
\begin{equation}
 \langle n^{(F)}_{k}\rangle=\langle{n}_{k,1}\rangle_{H=0}+\langle{n}_{k,2}\rangle_{H=0}
 =\frac{1}{e^{\beta(\varepsilon_k-\mu^{(F)}_{\infty})}+1},
 \label{eq:fsl-exc}
 \end{equation}
 where $\mu^{(F)}_{\infty}$ is the chemical potential of the free spinless Fermi gas:
\begin{equation}
\mu^{(F)}_{\infty}(T,n)=\mu_{\infty}+k_BT\ln{2}.  \label{eq:musl-exc}
\end{equation}

Lastly, using Eqs. (\ref{eq:fsl-exc}) and (\ref{eq:musl-exc}), the zero-field entropy per site in Eq. (\ref{peso}) 
can be written as
\begin{eqnarray}
 \frac{S_{\infty}(T,n)}{L}&=&nk_B \ln{2}-\frac{k_B }{L}\sum_k[\langle n^{(F)}_{k}\rangle \ln \langle n^{(F)}_{k}\rangle\nonumber\\
         & &+(1-\langle n^{(F)}_{k}\rangle)\ln(1-\langle n^{(F)}_{k}\rangle)]\label{sfermion1}\\
         &=&nk_B\ln{2}+\frac{S_{\infty}^{(F)}(T,n)}{L},\label{sfermion2}
\end{eqnarray}
where $S^{(F)}_{\infty}$ is the entropy of the free spinless Fermi gas. We stress that Eqs. (\ref{sfermion1})-(\ref{sfermion2}) or (\ref{peso}) in
zero field reproduce the two low-$T$ leading terms in Eq. (\ref{next}) in the limit 
$U=\infty$, i. e., $t^*=t$, after eliminating $\mu_{\infty}$ or $\mu^{(F)}_\infty$ 
in favor of $n$. Therefore, the specific heat calculated 
from either of the referred equations has the same value, since the difference between the 
two forms of the entropy function is a constant term, $nk_B\ln{2}$, associated with 
the disordered spin degrees of freedom.  

\section{Fractional Landau Luttinger liquid}
\label{sec:flll}
In the previous section,
we have described the low-energy physics of the Hubbard chain for $J(\equiv 4t^2/U)\ll k_BT\ll
E_F(\sim t)$ from the standpoint of a spin-incoherent LL, and have determined the
parameters $g$ and $v_c$ that govern this class of fluid. In this section, our aim is to show that the system can also be
mapped onto a fractional Landau LL \cite{prwu,242130,57977}.
This phenomenological approach, which is
a suitable generalization of the standard Landau Fermi liquid theory,
can shed light on the underlying aspects that characterize
the crossover behavior from the fixed point associated with $U=\infty$ at $T=0$ 
to the spin-incoherent LL regime at a given temperature $k_B T>J\ll t$.

\begin{figure}
\begin{center}
\includegraphics*[width=0.45\textwidth]{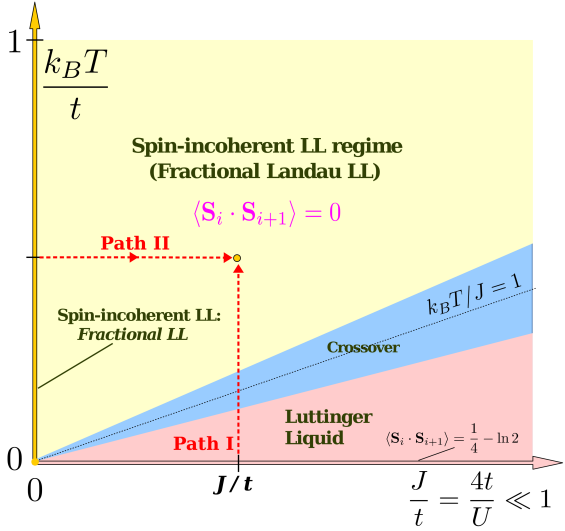}
\caption{(color online). Schematic phase diagram of the Hubbard chain in the strong-coupling regime, $J/t=4t/U\ll 1$, and  
$k_B T\lesssim t$. 
At the line $J=0$ the electrons are in a spin-incoherent Luttinger liquid (LL) phase with Curie response (
spin correlation length $\xi=0$). 
Further, this $U=\infty$ fixed point is exactly mapped 
onto an ideal gas with two species obeying the Haldane-Wu exclusion fractional statistics, i. e., a fractional LL. 
At the $T=0$ line, excluding the point $J=0$, the system is found in an LL phase
with algebraic decay of the charge and spin correlation functions. Increasing $T$ from a point at this line, Path I in the diagram, 
there is a crossover to a spin-incoherent regime with spin correlation length $\xi=v_s/(\pi k_BT)\sim [J/(2k_BT)] \ll 1$, so $\langle \mathbf{S}_i\cdot\mathbf{S}_{i+1}\rangle=0$. 
This regime can also be achieved through Path II, associated with both fractional LL and the fractional Landau LL: 
starting at $T=0$ and $U=\infty$, the temperature increases up to a value
at which the interaction is switched on and triggers the system into the spin-incoherent regime.
}
\label{fig:diag}
\end{center}
\end{figure}

In Fig. \ref{fig:diag} we present an schematic phase diagram $k_B T$ versus $J/t = 4t/U \ll 1$ that illustrates 
two possible thermodynamic paths
of the Hubbard model to
reach the spin-incoherent LL regime. The first one (Path I) is physically attained by increasing the temperature of the system, initially in the ground state of the
strong-coupling regime of the LL. The system undergoes a crossover and ends up at $T\gg J/k_B$, a spin disordered regime characterized by a zero pair
spin correlation function: $\langle {\bf S}_i\cdot {\bf S}_{i+1}\rangle=\gamma_S(T)=0$, as
discussed in Section \ref{sec:incohell}. In the second path (Path II), which helps us to
understand the Landau LL approach, the system starts at the fixed point $T = 0$ and $U = \infty$, the temperature increases up to a value
at which the interaction is switched on and triggers the system into the spin-incoherent regime. 

We thus assume that when corrections of ${\cal O}(t^2/U)$ are switched
on, the low-energy spectrum can be
obtained from the following expansion of the functional $E_L(T)-E_0(T=0)$:
\begin{eqnarray}
E_L(T)-E_0(T=0)&=&\sum_{k,\alpha}\tilde{\varepsilon}_{k,\alpha}\delta\langle\hat{n}_{k,\alpha}\rangle
\nonumber\\
&+&\frac{1}{2}\sum_{k,\alpha,k',\alpha'}f_{k,\alpha;k',\alpha'}\delta\langle\hat{n}_{k,\alpha}\rangle
\delta\langle\hat{n}_{k',\alpha'}\rangle,\nonumber\\
&&
\label{dog}
\end{eqnarray}
where $E_0(T=0)$ is the ground state energy,
\begin{equation}
\tilde{\varepsilon}_{k,\alpha}=-2t^{\ast}\cos{k},
\end{equation}
$t^{\ast}$ is the renormalized hopping amplitude with no effect of quasiparticle interaction,
\begin{equation}
\delta\langle\hat{n}_{k,\alpha}\rangle= \langle\hat{n}_{k,\alpha}(T)\rangle-
\langle\hat{n}_{k,\alpha}(0)\rangle,
\label{inferno}
\end{equation}
and $f_{k,\alpha;k',\alpha'}$
represents the interaction energy between quasiparticles. In addition, it is assumed
that the entropy has the same \textit{fractional} functional form of $S_{\infty}$,
Eq.~(\ref{peso}):
\begin{eqnarray}
&&S(T,\mu,H)=-k_B\sum_{k}[\langle{\hat n}_{k,1}\rangle\ln{\langle{\hat n}_{k,1}\rangle}
+\langle{\hat n}_{k,2}\rangle\ln{\langle{\hat n}_{k,2}\rangle}
\nonumber\\
 &&+(1-\langle{\hat n}_{k,1}\rangle-\langle{\hat n}_{k,2}\rangle)
\ln{(1-\langle{\hat n}_{k,1}\rangle-\langle{\hat n}_{k,2}\rangle)}].
\label{pesos}
\end{eqnarray}
It means that the statistics of the fractional
quasiparticles are also governed by the statistical matrix (\ref{canto}).

The equilibrium distribution of the quasiparticles is obtained by solving the equation
  $\partial\Omega/\partial\langle\hat{n}_{k,\alpha}\rangle=0$, where $\Omega=E-TS-\mu N$ and
 \begin{equation}
 n=\frac{1}{L}\sum_{k,\alpha}\langle\hat{n}_{k,\alpha}\rangle.
 \label{numero}
 \end{equation}
  After some algebra, one finds a distribution that is formally identical to Eq.~(\ref{mesa}):
 \begin{eqnarray}
\langle{\hat n}_{k,\alpha}\rangle=\frac{e^{-\beta({\hat\varepsilon}_{k,\alpha}-\mu_L)}}
{{\displaystyle 1+\sum_{\lambda=1}^{2}e^{-\beta({{\hat\varepsilon}}_{k,\lambda}-\mu_L)}}},
\label{mesas}
\end{eqnarray}
where
 \begin{equation}
 \hat{\varepsilon}_{k,\alpha}=\tilde{\varepsilon}_{k,\alpha}+
\sum_{k',\alpha'}f_{k,\alpha;k',\alpha'}\delta\langle\hat{n}_{k',\alpha'}\rangle
\label{pincelada}
\end{equation}
is the energy of the fractional Landau LL quasiparticle \cite{pines}.
 By symmetry considerations, the
 interaction energy between quasiparticles satisfies:
\begin{eqnarray}
f_{k,1;k',1}=f_{k,2;k',2}\equiv f_{k,k'}^{s}+f_{k,k'}^{a},\\
f_{k,2;k',1}=f_{k,1;k',2}\equiv f_{k,k'}^{s}-f_{k,k'}^{a},
\end{eqnarray}
which define the spin symmetric $f_{k,k'}^{s}$ and spin antisymmetric $f_{k,k'}^{a}$
parts of the fractional quasiparticle interaction \cite{pines}. In terms of these quantities, one has in zero field
\begin{equation}
\hat{\varepsilon}_{k,1}=\hat{\varepsilon}_{k,2}=-2t^{\ast}\cos{k}
+2\sum_{k'}f_{k,k'}^{s}\delta\langle\hat{n}_{k'}\rangle
\equiv \hat{\varepsilon}_{k}.
\end{equation}

In the following, it is our task to demonstrate that the above phenomenological approach proves useful in the understanding of the underlying
low-energy behavior of the Hubbard chain in the spin-incoherent regime. We emphasize that, regardless the fact that the
quasiparticles effects occur in the neighborhood of the Fermi surface $\{\pm k_F\}$, the final results are shown to be fully
compatible with those derived in the previous sections through a proper identification of the fractional Landau LL parameters.

 \subsection{Thermodynamic properties}

In order to compute the specific heat $C(T,n)$, we make the usual Landau assumption of neglecting
corrections to $\hat{\varepsilon}_{k,\alpha}$ due to interaction between the quasiparticles,
so that only the hopping amplitude is renormalized:
\begin{equation}
\hat{\varepsilon}_{k}\simeq \tilde{\varepsilon}_{k}
=-2t^{\ast}\cos{k}.
\label{tela}
\end{equation} 
Next, we insert Eq.~(\ref{tela}) into Eq.~(\ref{numero}) in order to obtain the fractional Landau LL chemical potential, $\mu_L$:
\begin{eqnarray}
\mu_L(T,n)&=&-2t^{\ast}\cos(n\pi)-k_BT\ln{2}
\nonumber\\ &+&\frac{\pi^2\cos(n\pi)(k_BT)^2}{12t^{\ast}\sin^2(n\pi)}+\cdots;
\label{sabonete}
\end{eqnarray}
therefore, the fractional Landau LL energy per site, and the fractional Landau LL specific heat, thus read:
\begin{eqnarray}
\frac{E_L(T,n)}{L}-\frac{E_0(T=0,n)}{L}&=&\frac{2}{L}\sum_{k}\tilde{\varepsilon}_{k}\left[\langle \tilde{n}_k\rangle
- \langle\hat{n}_{k}(0)\rangle\right]
\nonumber\\
&=&\frac{\pi(k_BT)^2}{12t^{\ast}\sin(n\pi)}+\cdots,
\label{ira}
\end{eqnarray}
where 
\begin{equation}
\langle \tilde{n}_k\rangle=\frac{1}{e^{\beta(\tilde{\varepsilon}_{k}-\mu_L)}+2}; \label{nlandau}
\end{equation}
and
\begin{equation}
C_L(T,n)=\frac{\pi k_B^2 T}{6t^{\ast}\sin(n\pi)} +\cdots.
\label{taxi}
\end{equation}

Comparing the above equation with Eqs. (\ref{za}), (\ref{sexta})-(\ref{girl}) associated with $C_{inch}$, we confirm 
our choice of $t^{\ast}$ in Eq. (\ref{za}).


The consistence of the fractional Landau LL approach is confirmed by the prediction for the entropy. In fact,
by using Eqs. (\ref{tela}),  (\ref{sabonete}) and (\ref{nlandau}) into Eq. (\ref{pesos}), we obtain
\begin{equation}
\frac{S_{L}(T,n)}{L}=nk_B\ln{2}+\frac{\pi k_B^2T}{6t^{\ast}\sin(n\pi)}+\cdots,
\label{entropyll}
\end{equation}
in complete agreement with $S_{inch}(T,n)$ in Eq.~(\ref{next}). 
Remarkably, the fractional Landau LL quasiparticles carry all the entropy of the system in the 
spin-incoherent regime $J\ll k_B T\ll E_F$, and correctly describe the fermionic spinless charge degrees 
of freedom and the background of fully disordered spin degrees of freedom.

The prediction for $\kappa$ is obtained as follows. From $n=\sum_{k}2\langle\hat{n}_k\rangle/L$, we get
\begin{equation}
\frac{\partial n}{\partial\mu}=\frac{2}{L}\sum_{k}\frac{\beta (1-\partial{\hat{\varepsilon}_k}/\partial\mu)
e^{\beta(\hat{\varepsilon}_k-\mu)}}{[e^{\beta(\hat{\varepsilon}_k-\mu)}+2]^2},
\label{verdureiro}
\end{equation}
where
\begin{equation}
\frac{\partial{\hat{\varepsilon}_k}}{\partial\mu}=2\sum_{k'}\frac{f_{k,k'}^{s}\beta
(1-\partial{\hat{\varepsilon}_{k'}}/\partial\mu) e^{\beta(\hat{\varepsilon}_{k'}-\mu)}}{[e^{\beta(\hat{\varepsilon}_{k'}-\mu)}+2]^2}.
\label{verdura}
\end{equation}
At low-$T$, the above integrands have sharp peaks centered at the $k$ vectors
 of the Fermi surface $\{\pm k_{F}\}$; therefore, one obtains (see Appendix \ref{tudo})
\begin{eqnarray}
\frac{\partial{\hat{\varepsilon}_k}}{\partial\mu}&=&(f_{k,k_F}^{s}+f_{k,-k_F}^{s})
 \sum_{k'}\frac{\beta (1-\partial{\hat{\varepsilon}_{k'}}/\partial\mu)
 e^{\beta(\hat{\varepsilon}_{k'}-\mu)}}{[e^{\beta(\hat{\varepsilon}_{k'}-\mu)}+2]^2}
\nonumber\\
&=&(f_{k,k_F}^{s}+f_{k,-k_F}^{s})\left(\frac{L}{2}\right)\left(\frac{\partial n}{\partial\mu}\right).
\label{fruta}
\end{eqnarray}
 By inserting this back into $\frac{\partial n}{\partial\mu}$ and using $\kappa^{-1}=n^2(\partial \mu / \partial n)$, we find
\begin{equation}
\kappa_L^{-1}(T,n)=2\pi t^{\ast}n^2\sin(n\pi)(1+F_0^{s})+\cdots,
\label{patinha}
\end{equation}
where
\begin{equation}
F_0^{s}=\frac{L(f_{k_F,k_F}^{s}+f_{k_F,-k_F}^{s})}{4\pi t^{\ast}\sin(n\pi)}
\label{zaza}
\end{equation}
 is the Landau-Luttinger parameter associated with the spin symmetric part of the
 quasiparticle interaction at the Fermi level $(k_F=n\pi)$.
 A comparison of Eqs.~(\ref{patinha}) and (\ref{indio}) implies:
\begin{equation}
 F_0^s= -\frac{v_{c,\infty}}{\pi U},
 \end{equation}
with $v_{c,\infty}=v_c^{(inch)}|_{U=\infty}=2t\sin(n\pi)$.
Notice that $F_0^s$ is in fact the ratio of the
total kinetic energy per site for $U=\infty$ at $T=0$ over the
on-site Coulomb repulsion $U$.

We now calculate the prediction for $\chi$. In the presence of a magnetic field, we replace
$\hat{\varepsilon}_{k}$ by $\hat{\varepsilon}_{k,\alpha}=\tilde{\varepsilon}_{k}\mp \mu_BH$ in Eq.~(\ref{mesas}).
Thus the spin susceptibility is given by
\begin{eqnarray}
&&\chi_L(T,n)=\frac{\mu_B^2}{L}\sum_{k}
\frac{\partial}{\partial H}(\langle\hat{n}_{k,1}
\rangle - \langle\hat{n}_{k,2}\rangle)_{H=0}
\nonumber\\
&&=\frac{ \mu_B^2}{L}\sum_{k}\frac{\beta}{[e^{\beta(\hat{\varepsilon}_k-\mu)}+2]}
 \frac{\partial}{\partial H}( \hat{\varepsilon}_{k,2}- \hat{\varepsilon}_{k,1})_{H=0},
\nonumber\\
\label{grampo}
\end{eqnarray}
where
\begin{eqnarray}
\frac{\partial}{\partial H} (\hat{\varepsilon}_{k,2} - \hat{\varepsilon}_{k,1} )_{H=0}=2
\nonumber\\
+2\sum_{k'}f^{a}_{k,k'} \frac{\partial}{\partial H} (\langle\hat{n}_{k',2} \rangle-
 \langle\hat{n}_{k',1} \rangle )_{H=0}.
 \end{eqnarray}
Since we expect $f_{k,k'}^{a}={\cal O}(t^2/U)$, we can take
\begin{equation}
 \frac{\partial}{\partial H}(\langle\hat{n}_{k',2} \rangle -
 \langle\hat{n}_{k',1} \rangle)_{H=0}=-\frac{2\beta}{e^{\beta(\hat{\varepsilon}_{k'}-\mu)}+2}
 \end{equation}
in the last expression. Therefore, the spin susceptibility becomes
\begin{equation}
\chi_L(T,n)=\frac{\mu_B^2n}{k_BT}(1-\beta t F_0^{a})+\cdots,
\label{suco}
\end{equation}
where
\begin{equation}
F_{0}^{a}=\frac{4}{tN}\sum_{k}\sum_{k'}f_{k,k'}^{a}\langle \hat{n}_{k}\rangle
\langle \hat{n}_{k'}\rangle.
\label{ter}
\end{equation}
In contrast to Eqs.~(\ref{verdureiro}) and (\ref{verdura}), the
absence of sharp peaks at the Fermi surface in Eq.~(\ref{ter}) is a clear
manifestation of the fact that the spin degrees of freedom are highly thermalized.
 A comparison between Eqs.~(\ref{graviola}) and (\ref{suco}), however, allows us to identify
\begin{equation}
F_{0}^{a}=\frac{nv_s}{\pi t}
\label{foton}
\end{equation}
without the need of specifying the range of integration.
 If, in addition, we make the assumption that $f_{k,k'}^{a}$ is $k$-independent, Eq.~(\ref{foton}) implies $Lf^{a}=v_s/\pi$. Notice also that $F_{0}^{a}$ is the ratio between the energy per site of the
 Heisenberg Hamiltonian in the spin-incoherent regime and $nt$ [see Eqs.~(\ref{pasta}), (\ref{pastel})-(\ref{GSenergy}) and (\ref{gammaS}) ].

Lastly, we shall digress on the eventual crossover of the magnetic susceptibility as $T\rightarrow 0$.
Unlike the crossover associated with the charge response functions, which is governed by 
the spin-spin correlation function, as discussed in Sec. \ref{sec:incohell}, in 
the magnetic susceptibility case there is a change of paradigm as $T\rightarrow 0$.
First, as a guess, we notice that, to ${\cal O}(t^2/U)$:
$\lim_{T\rightarrow 0}\chi_L(T,n)=\lim_{T\rightarrow 0} \frac{\mu_B^2\beta n}{1+\beta t F_0^a}=\pi\mu_B^2/v_s$.
It thus suggests the following \textit{ansatz} for the Landau
parametrization: $\lim_{T\rightarrow 0}(tF_0^a)=(-\beta^{-1}+n \pi v_s/2)$, 
which implies \cite{essler2005one} $\lim_{T\rightarrow 0}\chi_L(T,n)=2\mu_B^2/\pi v_s$. It entails that, as $T\rightarrow 0$, the strong-coupling exchange enhancement of ${\cal O}(t^2/U)$ suppresses the Curie behavior and gives rise to 
the LL power-law decay of the spin correlation function and the very low-$T$ behavior of $C(T)$ shown in Fig. \ref{cvfiete}, 
with dominant spinon contribution, see Eq. (\ref{gammat0}).

 \subsection{Drude Weight}

In the presence of an external electric field $\phi$, the spectrum $E_{\infty}$ of the Hubbard chain
with $U=\infty$, or $J=0$ in Eq. (\ref{eq:tj}), 
is altered according to the well known prescription \cite{65243,*Kohn1964}
\begin{equation}
E_{\infty}\rightarrow \sum_{k}\varepsilon_k(\phi)n_k,
\end{equation}
 where
 \begin{equation}
 \varepsilon_k(\phi)=-2t\cos(k+\phi).
 \end{equation}
Since Eqs.~(\ref{dog}) and (\ref{pincelada}) establish an one-to-one mapping
between the eigenstates of the Hamiltonian for $J=0$ and
$J\neq0$, in the presence of $\phi$ we have
 \begin{eqnarray}
E(\phi)-E_0=\sum_{k,\alpha}\tilde{\varepsilon}_{k,\alpha}(\phi) \delta\langle\hat{n}_{k,\alpha}\rangle_{\phi}
\nonumber\\
+\frac{1}{2}\sum_{k,\alpha,k',\alpha'}f_{k,\alpha;k',\alpha'}\delta\langle\hat{n}_{k,\alpha}\rangle_{\phi}
\delta\langle\hat{n}_{k',\alpha'}\rangle_{\phi},
\label{amendoim}
\end{eqnarray}
where
\begin{eqnarray}
&&\tilde{\varepsilon}_{k,\alpha}(\phi) =-2t^{\ast}\cos(k+\phi),
 \\
&&\hat{\varepsilon}_{k,\alpha}(\phi)=\tilde{\varepsilon}_{k,\alpha}(\phi)+\sum_{k', \alpha'}f_{k,\alpha; k',\alpha'}
\delta\langle\hat{n}_{k',\alpha'}\rangle_{\phi},
\label{terreno}
\\
&&\delta\langle\hat{n}_{k,\alpha}\rangle_{\phi}=\frac{1}{e^{\beta[\hat{\varepsilon}_{k,\alpha}(\phi)-\mu]}+2}
-\langle\hat{n}_{k,\alpha}\rangle_{T=0\atop\phi=0}.
\label{sacola}
\end{eqnarray}
We are now in a position to obtain the Drude weight \cite{65243,*Kohn1964} (see Appendix \ref{cru}):
\begin{eqnarray}
\sigma_0&=&
-\frac{\pi}{L}\left[ \frac{\partial^2 E(\phi)}{\partial\phi^2} \right]_{\phi=0}
\nonumber\\
&=&2t^{\ast} \sin(n\pi)-\left(\frac{L}{\pi}\right)(f_{k_F,k_F}^{s}-f_{k_F,-k_F}^{s}).
\label{liga}
\end{eqnarray}
Now using Eq.~(\ref{trans}), one obtains
\begin{equation}
 \frac{L(f_{k_F,k_F}^{s}-f_{k_F,-k_F}^{s})}{2\pi t^{\ast}\sin(n\pi)}=F_{0}^{s}.
 \label{oito}
 \end{equation}
 A combination of Eqs.~(\ref{zaza}) and (\ref{oito}) determines the
 spin symmetric part of the interaction energy between quasiparticles:
 \begin{eqnarray}
 L f_{k_F,k_F}^{s}=\frac{3\pi v_{c,\infty}}{2}(1-1/g),
 \label{ca}\\
 L f_{k_F,-k_F}^{s}=\frac{\pi v_{c,\infty}}{2}(1-1/g),
 \label{cb}
 \end{eqnarray}
 with $v_{c,\infty}=v_c^{(inch)}|_{U=\infty}$.
 Note in addition that the renormalized hopping can be written as
 \begin{equation}
 t^{\ast}=t\left[\frac{v_c^{(inch)}}{v_{c,\infty}}\right].
 \label{cc}
 \end{equation}
It is now clear that Eq.~(\ref{foton}) and Eqs.~(\ref{ca})-(\ref{cc}) establish the
connection between the fractional Landau LL
parametrization and that of the LL in the spin-incoherent regime.
\subsection{Specific heat and numerical data}
We shall now demonstrate 
that in the spin-incoherent regime the fractional Landau LL approach provides
a very good description of the $T$-behavior of the zero-field specific heat of the system
derived from the entropy defined in Eq. (\ref{pesos}).
We stress that this procedure will prove 
rewarding in establishing an exact connection between the fractional Landau LL and an \textit{interacting} spinless Fermi gas, 
similarly to the 
one that we have discussed between the fractional LL and the \textit{free} spinless Fermi gas in Section \ref{sec:uinfity}. However, 
the fractional Landau LL is valid only under the condition $J (=4t^2/U) \lesssim k_BT \lesssim E_F (\sim t)$, while the fractional LL is an exact description
at $U=\infty$ and any $T$.

In zero field, using the Landau assumption in the calculation
of the specific heat, Eqs. (\ref{tela}) and (\ref{nlandau}), 
and summing up the two fractional species we can obtain a direct relation
between $\langle \tilde{n}_k \rangle$ and the \textit{interacting} spinless
Fermi gas distribution function:

\begin{equation}
 \langle \tilde{n}^{(F)}_{k}\rangle=2\langle \tilde{n}_k\rangle=\frac{1}{e^{\beta(\tilde{\varepsilon}_k-{\mu}^{(F)}_L)}+1},
\label{eq:nk-fll}
 \end{equation}
with
\begin{equation}
\mu^{(F)}_L=\mu_L+k_BT\ln2,
\label{eq:mu-f-e}
\end{equation}
where $\mu_L$ is the fractional Landau LL chemical potential and $\mu^{(F)}_L$ is the chemical potential of the related 
interacting spinless Fermi gas. 
Lastly, by replacing $\langle \hat{n}_k\rangle \rightarrow \langle \tilde{n}_k\rangle$ in Eq. (\ref{pesos}), and 
using Eqs. (\ref{eq:nk-fll}) and (\ref{eq:mu-f-e}), we can obtain a relation 
between the fractional Landau LL entropy, $S_{L}$, and the related interacting spinless Fermi gas entropy, $S_{L}^{(F)}$: 
\begin{eqnarray}
 \frac{S_{L}(T,n)}{L}&=&nk_B \ln{2}-\frac{k_B }{L}\sum_k[\langle \tilde{n}^{(F)}_{k}\rangle \ln \langle \tilde{n}^{(F)}_{k}\rangle\nonumber\\
         & &+(1-\langle \tilde{n}^{(F)}_{k}\rangle)\ln(1-\langle \tilde{n}^{(F)}_{k}\rangle)]\label{eq:sl-f}\\
         &=&nk_B\ln{2}+\frac{S_{L}^{(F)}(T,n)}{L},
\end{eqnarray}
which is formally identical to Eq. (\ref{sfermion2}) at $U=\infty$.
\begin{figure}
\begin{center}
\includegraphics*[width=0.45\textwidth]{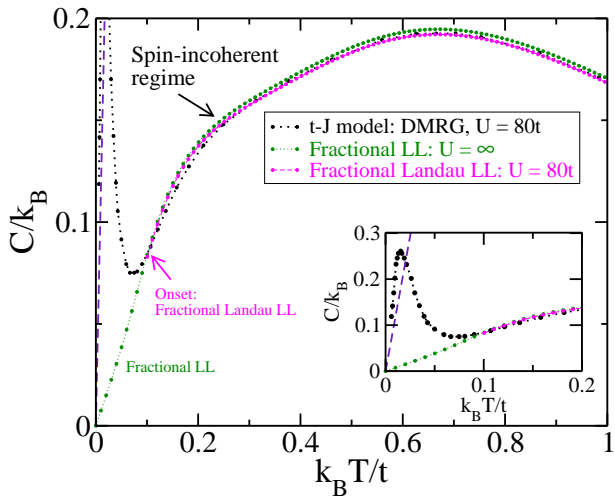}
 \caption{(color online). Specific heat $C$ in units of $k_B$ as a function of the thermal energy $k_B T$ in units of $t$ for chains with $n=3/4$. 
 The DMRG data for a $t-J$ chain with 32 sites and $J=0.05t$ ($U=80t$), from Ref. \onlinecite{81075108}. 
 Also shown are predictions from the fractional Landau LL in zero field  
 for $U=80t$ and the fractional LL at $U=\infty$. 
 Notably, the results of the fractional Landau LL are in very good agreement with
 the DMRG data in the spin-incoherent regime. 
 For completeness, we show the straight line of the $T\rightarrow 0$ limit of $C/k_B$, whose coefficient is $\gamma k_BT$, with $\gamma$ in Eq. (\ref{gammat0}).
 The insert shows details of the referred estimates 
 for $C/k_B$ in a narrow low $T$-interval. Notice that in Fig. \ref{fig:diag} Path I is associated with the DMRG data, while Path II
 with the fractional LL and the fractional Landau LL.}
\label{cvfiete}
\end{center}
\end{figure}

The function $\mu_L(T,n)$, to order $(k_B T/t^\ast)^2$, is given by
Eq. (\ref{sabonete}); however, in order to attain a good description for a wide range of temperatures 
we have calculated $\mu_L(T,n)$ numerically
using the constraint equation
\begin{equation}
 \frac{2}{L}\sum_{k}\langle \tilde{n}_k\rangle = n,
 \label{eq:vinc-exc}
\end{equation}
where $n$ is the average density of spinless fermions.

From either entropy above, we can numerically calculate the specific heat of the fractional Landau LL gas using $C=T(\partial S/\partial T)$.
In Fig. \ref{cvfiete} we show $C(T)/k_B$ for the fractional LL ($U=\infty$) and the fractional Landau LL for $U=80t$ ($J=0.05t$)
for chains with $n=3/4$. 
The specific heat of the fractional Landau LL in zero field is derived using Eqs. (\ref{pesos}), (\ref{tela}) and (\ref{za}), 
 for $U=80t$, whereas for the fractional LL, $U=\infty$, use is made of Eq. (\ref{peso}).
Remarkably, the fractional Landau LL prediction quantitatively agrees with the 
DMRG data in the temperature range of the spin-incoherent regime up to $k_B T\sim t$. Despite the tiny value of $\frac{t}{t*}=0.987$, 
the fractional Landau LL approach adequately quantifies the first order correction, $(t/U)$, to the $U=\infty$ curve 
in the spin-incoherent regime.

The two paths to the spin-incoherent LL regime shown in Fig. \ref{fig:diag} can be discussed 
with the aid of Fig. \ref{cvfiete}. The Path I of Fig. \ref{fig:diag} is associated with the DMRG data of Ref. \onlinecite{81075108} 
showed in Fig. \ref{cvfiete}, in which case we witness
the linear behavior of the specific heat, with spin and charge contributions at very low temperature, and the crossover to
the spin-incoherent regime. Further, Path II of Fig. \ref{fig:diag} is associated with the analytical results plotted 
in Fig. \ref{cvfiete}. Indeed, in this figure we indicate the onset of the spin-incoherent regime, in which
case we can notice that the specific heat data of the fractional LL and that of the Landau fractional LL, both due
to charge contribution only, practically meet at the onset of the spin-incoherent regime, since they differ by the
small correction term of order $t/U$.

\section{High-temperature limit}
\label{sec:hight}
\begin{figure}
\begin{center}
\includegraphics*[width=0.4\textwidth]{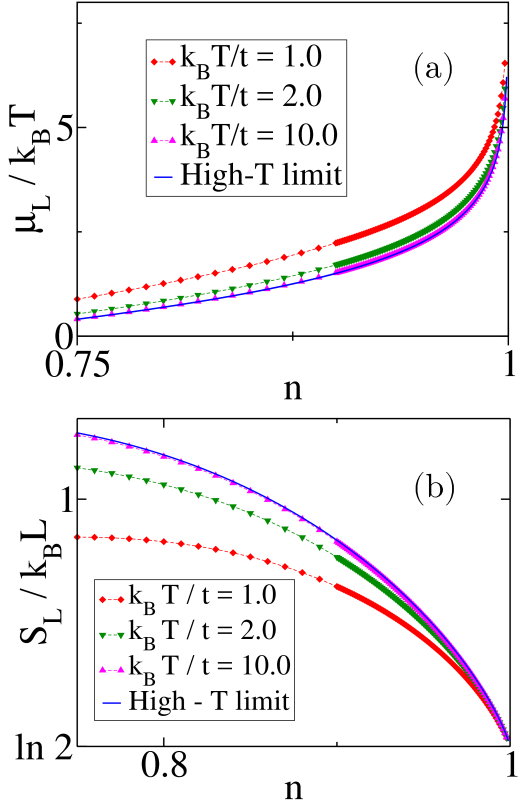}
\caption{(color online). (a) Fractional Landau LL chemical potential, $\mu_L$, in units of $k_BT$ as a function of 
 particle density $n$, for $U=80t$. The high-$T$ limit, given by
 $\frac{\mu_L}{k_BT}=\ln(\frac{n}{2-2n} )$, is indicated. (b) Fractional Landau LL entropy per site, $S_L/L$, in 
 units of $k_B$, as a function of $n$, for $U=80t$. The high-$T$ limit, given by
 $\frac{S_L}{k_BL}=n\ln{2}-n\ln{n}-(1-n)\ln{(1-n)}$, is also shown.}
\label{fig:semuhight}
\end{center}
\end{figure}

In previous sections we have studied the Hubbard chain in the spin-incoherent regime: $J\ll k_BT \ll t$, 
using a perturbative Bethe ansatz procedure, valid for $U/k_B T\gg 1$, combined with 
a phenomenological approach. In this Section, we find it instructive to study the high-$T$ limit, so we can provide direct contact with
well established results for the $t$-$J$ models derived using quantum transfer matrix techniques \cite{Juttner1997}.
The high-$T$ limit is accessed under the conditions: $e^{-\beta \tilde{\varepsilon}_k}\rightarrow 1$, 
with $\frac{\mu_L}{k_BT}$ a function of $n$. Indeed, from either Eqs. (\ref{eq:nk-fll})-(\ref{eq:mu-f-e}) or Eq. (\ref{eq:vinc-exc}),
 we find that $\langle \tilde{n}_k\rangle=n/2$ and
\begin{equation}
 \lim_{T\rightarrow\infty}\frac{\mu_L}{k_BT}=\ln\left(\frac{n/2}{1-n}\right),
 \label{eq:htmul}
\end{equation}
which exhibits a Van-Hove singularity as $n\rightarrow 1$, as illustrated in Fig. \ref{fig:semuhight}(a).
These results imply that $S_L$ in Eq. (\ref{eq:sl-f}) reads:
\begin{equation}
\lim_{T\rightarrow\infty}\frac{S_L(T,n)}{k_BL}=n\ln{2}-n\ln{n}-(1-n)\ln{(1-n)},
\label{eq:hts}
\end{equation}
which is exactly the result expected by counting the total number of states of 
the $t-J$ model at a density $n$, with $N_\uparrow = N_\downarrow$, in the thermodynamic limit.
In Fig. \ref{fig:semuhight}(b) we present $S_L(T,n)$ with a density $n$ for $U=80t$. 
Its also interesting to notice that $S_L(T,n)/k_B L$ approaches $\ln 2$ at half-filling due 
to the Van-Hove singularity. 
In addition, we stress that the high-$T$ limit is taken under the proviso that 
$U/k_B T\gg 1$, as is the case in Figs. \ref{fig:semuhight}(a) and (b), in which case 
$U=80t$. It is worth mentioning that as $T\rightarrow\infty$, $U$ increases accordingly, so that,
Eq. (\ref{eq:hts}) is the $T\rightarrow\infty$ entropy of the $U=\infty$ Hubbard chain, Eq. (\ref{sfermion2}).

Lastly, in order to confirm the high-$T$ limit of the particle occupation number, $\langle n_k\rangle$, of the $t$-$J$ model, Eq. (\ref{eq:tj}),
we use the Lanczos exact diagonalization and finite temperature Lanczos method (FTLM) \cite{ftlm} 
to calculate $\langle n_k\rangle$ in finite chains under periodic boundary conditions (PBC).
In fact, our analysis provides strong evidence in favor of our 
analytical results and, most importantly, verifies the consistency of the fractional Landau LL phenomenological 
approach. 


The FTLM uses the states from $R$ independent Lanczos exact diagonalization procedures 
to estimate thermodynamic functions of finite systems.
For each Lanczos run, a maximum of $M$ Lanczos basis states is generated. 
The $MR$ approximate eigenenergies and eigenstates are used to calculate the thermodynamic functions
of interest. We take $R=12000$ and $M=50$ in our calculations, and have exploited translational symmetry and rotational
symmetry in spin space.
\begin{figure}
\begin{center}
\includegraphics*[width=0.45\textwidth]{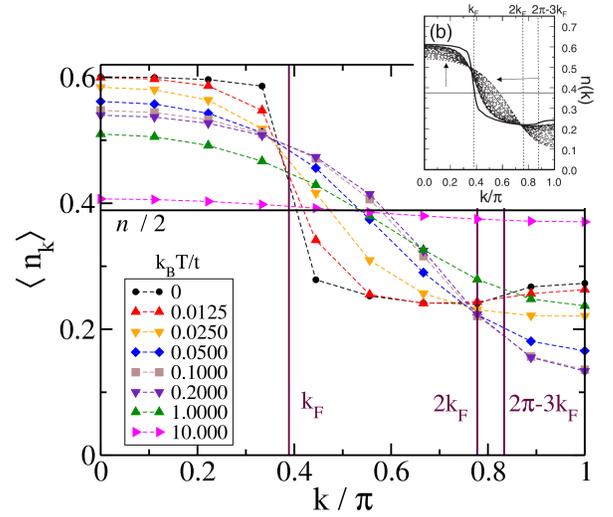}
\caption{(color online). Momentum distribution function $\langle n_k\rangle$ calculated 
through FTLM for a $t-J$ chain with 18 sites under PBC, particle density $n=7/9$, $J=0.05t$, and $k_F=n\pi/2$ 
for the indicated values of temperature $k_B T/t$. 
Notice that the limit $\langle n_k\rangle=n/2$ as $(k_B T/t)\rightarrow\infty$ is nearly attained for $(k_B T /t)=10$.
The inset is a copy of Fig. 3(b) of Ref. \onlinecite{81075108}:  DMRG data 
for $n=0.75$, $J=0.05t$, for a chain with 32 sites. Arrows indicate increasing $t/k_B T$ in steps of 4.
The horizontal line at $\langle n_k\rangle=n/2=0.375$ indicates the value of $\langle n_k\rangle$ for $(k_B T/t)\rightarrow\infty$.}
\label{nk}
\end{center}
\end{figure}

The distribution function of spin $\uparrow$ electrons of momentum $k$ is calculated through:
\begin{equation}
 \langle n_k\rangle =\frac{1}{L}\sum^L_{l=1,m=1} \langle c^{\dagger}_{l\uparrow} c_{m\uparrow}\rangle e^{ik(l-m)},
\end{equation}
where $\langle \ldots \rangle$ indicates thermal and quantum averages.
In Fig. \ref{nk} we present $\langle n_k\rangle$ for $J=0.05t$ and $n=7/9$, calculated with the Lanczos method ($T=0$) and FTLM ($T\neq 0$), 
as well as DMRG data from Ref. \onlinecite{81075108} 
for $n=0.75$ and $J=0.05t$, shown in the inset.
At $T=0$, the singularities \cite{412326,81075108} at $k_F$ and $3k_F$ (shown at $2\pi-3k_F$) are evident in our FTLM results for $(k_BT/t)=0$ and $0.0125$,
with $k_F=\pi n/2$. 
The spin-incoherent regime, $k_B T\gtrsim J$, is signaled \cite{81075108} by the presence of an inflection point at $2k_F$, 
as observed in Fig. \ref{nk} for $(k_B T/t) =0.05$, 0.10 and 0.20. 
We thus conclude that both the FTLM and DMRG methods grasp the main features of the crossover between the low-$T$ LL to its spin-incoherent
regime.

\section{Concluding Remarks}
\label{sec:remarks}
We have studied the Hubbard chain in the spin-incoherent Luttinger liquid regime, both for $J=0$ and finite $J(\ll k_BT)$. In the former
case, we have shown that its thermodynamic properties are exactly those of an ideal gas of two species of noninteracting particles obeying fractional statistics.
It implies that the charge degrees of freedom are governed by the free spinless Fermi gas, while the spin degrees of freedom are fully
disordered (Curie response). On the other hand, the latter case was investigated using an expression for the grand-canonical free
energy derived  perturbatively by Ha from Takahashi's
integral equations. Based on this result, and using $U\gg k_B T$, we were able to obtain an expression for the Helmholtz free energy suitable to describe the system in the 
spin-incoherent regime: $J(\equiv 4t^2/U)\ll k_B T\ll E_F$,
from which  several thermodynamic quantities were derived. In particular, we have reported on the specific heat, charge compressibility, 
magnetic susceptibility, Drude weight, charge
and spin velocities, and the Luttinger liquid (LL) parameter. 

We have also discussed the interesting possibility of looking at the system with finite $J$ as a fractional Landau LL. In this framework, 
the low-energy physics of the system is also described in terms of fractional quasiparticles obeying 
the Haldane-Wu fractional entropy. 
At the same time, it enables us to interpret
corrections of ${\cal O}(t^2/U)$ as coming
from (i) renormalization of the hopping $t$ only, which is the case of the specific heat and charge velocity;
(ii) hopping renormalization and the interaction of fractional Landau quasiparticles, as found for the charge compressibility, and the Drude weight;
(iii) interaction of fractional Landau LL quasiparticles only, as for the magnetic susceptibility. In addition, we have calculated the fractional
Landau LL parameters and showed that they are fixed by the ones of the incoherent
LL derived from pure thermodynamic grounds and arguments of bosonization.
In particular, a phase diagram was provided and two thermodynamic paths to access the spin-incoherent LL 
regime shed light on the numerical and analytical procedures.
Lastly, through a numerical analysis of the excluson fractional Landau LL entropy and the use of finite temperature Lanczos method,
we have calculated the temperature behavior of the specific heat and 
the particle momentum distribution, respectively, both in very good agreement with previous
density matrix renormalization 
group calculations in the spin-incoherent and the high-$T$ limit.

In conclusion, we believe that our reported results using complementary approaches have provided interesting insights on several features of the 
thermodynamics of the spin-incoherent Luttinger liquid regime of the Hubbard chain.
They might stimulate further theoretical and experimental work, since this special LL regime 
has been of interest in the context of several physical systems mentioned in our work, 
particularly quantum wires at low temperature. In addition, the crossover \cite{57977,PhysRevB.61.7909,*PhysRevLett.87.276405,*Kung2017} 
from the (1D) spin-incoherent LL
regime (Fractional Landau LL) to a higher dimensional phenomenology (due to 2D or 3D coupling between chains), e. g., standard
Landau Fermi liquid theory, also deserves further investigation.

\section*{Acknowledgments}
We acknowledge financial support from Coordena\c{c}\~ao de
Aperfei\c{c}oamento de Pessoal de N\'{\i}vel Superior (CAPES),
Conselho Nacional de Desenvolvimento Cient\'{\i}fico e
Tecnol\'ogico (CNPq), and Funda\c{c}\~ao de Amparo \`a Ci\^encia e
Tecnologia do Estado de Pernambuco (FACEPE), Brazilian
agencies, including the PRONEX Program which is funded by
CNPq and FACEPE.

\appendix

\section{The grand-canonical free energy for $U\gg k_B T$, Eq. (\ref{tanto})}
\label{strHa}

In Ref. \onlinecite{46a}, Ha derived a strong coupling ($U\gg t$) 
perturbative $\lambda$-expansion of
the grand-canonical free energy $\frac{\Omega(T,\mu,H)}{L}$:

\begin{equation}
\frac{\Omega(T,\mu,H)}{L}=\omega^{(0)}+\omega^{(1)}+\ldots,
\label{a1}
\end{equation}
where
\begin{equation}
\omega^{(0)}=\frac{U}{2}-\mu-\frac{1}{\beta}\ln{2a}-\frac{1}{\beta}I_1
\end{equation}
and
\begin{equation}
\omega^{(1)}=\frac{t}{U}I_2\left[\left(\frac{1}{a^2}-\frac{1}{b^2}\right)\frac{1}{\beta}I_3-\frac{1}{a^2}\right];
\end{equation}
with
\begin{equation}
a=\cosh\left[\beta\left(\frac{U}{2}-\mu\right)\right],
\end{equation}
\begin{equation}
b=\cosh\left(\beta \mu_B H\right),
\end{equation}
\begin{equation}
I_1=\int^{\pi}_{-\pi}\frac{dk}{2\pi}\ln \left[1+\frac{b}{a} e^{-\beta\left(\varepsilon_k-\frac{U}{2}\right)}\right],
\end{equation}
\begin{equation}
I_2=\int^{\pi}_{-\pi}\frac{dk}{2\pi}\left[1+\frac{a}{b}e^{\beta\left(\varepsilon_k-\frac{U}{2}\right)}\right]^{-1},
\end{equation}
and
\begin{equation}
I_3=\int^{\pi}_{-\pi}\frac{dk}{2\pi}\cos k \ln \left[1+\frac{b}{a} e^{-\beta\left(\varepsilon_k-\frac{U}{2}\right)}\right],
\end{equation}
in which $\varepsilon_k=-2t\cos{k}$.

The above expansion was used to obtain two 
expansions in distinct limits:
(i) $U,~k_B T\gg t$ with $U/k_B T$ fixed, 
which was shown to be in very good agreement 
with previous high-$T$ expansions \cite{*[{See, e. g., }] [{}] Kubo1980}; (ii) $U\gg t$ at 
fixed $k_B T$. We shall use the latter alternative 
in the limit $U\gg k_B T$, in which case we have $e^{\beta U}\gg 1$ and $a\sim \frac{1}{2}e^{\beta\left(\frac{U}{2}-\mu\right)}\gg 1$, such that 
\begin{equation}
\omega^{(0)}=-k_BTI_1
\label{w0}
\end{equation}
and 
\begin{equation}
\omega^{(1)}=\frac{-t}{U}I_2\frac{1}{b^2}\frac{1}{\beta}I_3=\frac{-k_BT}{\cosh^2(\beta \mu_B H)}\left(\frac{t}{U}\right)I_2I_3;
\label{w1}
\end{equation}
with
\begin{equation}
I_1=\int^{\pi}_{-\pi}\frac{dk}{2\pi}\ln{\left[1+e^{-\beta(\varepsilon_k-\mu-\mu_BH)}+e^{-\beta(\varepsilon_k-\mu+\mu_BH)}\right]},
\label{i1}
\end{equation}
\begin{equation}
I_2=\cosh(\beta\mu_B H)\int^{\pi}_{-\pi}\frac{dk}{2\pi}\frac{2}{e^{\beta(\varepsilon_k-\mu)}+2\cosh(\beta\mu_B H)},
\label{i2}
\end{equation}
and
\begin{eqnarray}
I_3=\int^{\pi}_{-\pi}\frac{dk}{2\pi}\cos{k}\ln[1+e^{-\beta(\varepsilon_k-\mu-\mu_BH)}\nonumber\\ 
+e^{-\beta(\varepsilon_k-\mu+\mu_BH)}].
\label{i3}
\end{eqnarray}

Therefore, Eq. (\ref{a1}) with $\omega^{(0)}$ and $\omega^{(1)}$ given by Eqs. (\ref{w0}) and (\ref{w1}), 
with $I_1$, $I_2$, and $I_3$ defined through Eqs. (\ref{i1}), (\ref{i2}) and (\ref{i3}), respectively,
leads to Eq. (\ref{tanto}).
\section{Susceptibility at $H=0$, Eq. (\ref{graviola})}
\label{susInch}

The magnetic susceptibility per site for 
$H=0$ is given by 
\begin{equation}
\chi(T,\mu)=\left .\frac{\partial M(T,H,\mu)}{\partial H}\right |_{H=0}=-\left .\frac{\mu_B}{L}\frac{\partial^2 \Omega(T,H,\mu)}{\partial H^2}\right |_{\substack{H=0}},
\label{chiap}
\end{equation}
with $\Omega/L$ as written in Eq. (\ref{tanto}).

Using Eq. (\ref{chiap}), and taking the limits $\sinh(\beta\mu_B H)\rightarrow \beta \mu_B H$, 
$\tanh(\beta\mu_B H)\rightarrow \beta \mu_B H$, and $\cosh(\beta\mu_B H)\rightarrow 1$, 
we obtain the following 
expression for the susceptibility: 
\begin{equation}
  \chi=\frac{\mu_B^2}{k_BT}\left(I_{2,0}-\frac{t}{U}I_{2,0}I_{3,0}-\frac{4t}{U}I_{4,0}I_{3,0}+\frac{4t}{U}\frac{I_{2,0}}{2}I_{5,0}\right )
\label{chiii}
\end{equation}
where $I_{1,0}$, $I_{2,0}$, and $I_{3,0}$ are given by Eqs. (\ref{i1}), (\ref{i2}) and (\ref{i3}), respectively, 
with $H=0$, while
\begin{equation}
 I_{4,0}=\int_{-\pi}^{\pi}\frac{dk}{2\pi}\frac{1}{[e^{\beta(\varepsilon_k-\mu)}+2]^2},
\end{equation}
\begin{equation}
 I_{5,0}=\int_{-\pi}^{\pi}\frac{dk}{2\pi}\frac{\cos{k}}{e^{\beta(\varepsilon_k-\mu)}+2}.
\end{equation}
Now, by performing a Sommerfeld-like expansion around $\mu_{inch}(T=0)\sim t$, Eq. (\ref{muinch}), in the above integrals, 
we obtain 
\begin{equation}
 I_{2,0}=n+\ldots,
 \label{ii2}
\end{equation}
\begin{equation}
 I_{3,0}=\frac{nt}{k_BT}\left(1-\frac{\sin{2\pi n}}{2\pi n}\right)+\ldots,
 \label{ii3}
\end{equation}
\begin{equation}
 I_{4,0}=\frac{n}{4}+\ldots,
 \label{ii4}
\end{equation}
\begin{equation}
 I_{5,0}=\frac{\sin(n\pi)}{\pi}+\ldots .
\label{ii5}
\end{equation}

Lastly, by substituting Eqs. (\ref{ii2})-(\ref{ii3}) 
into Eq. (\ref{chiii}), we obtain
\begin{equation}
 \chi=\frac{\mu_B^2}{k_BT}\left[n-\frac{2n^2t^2}{Uk_BT}\left(1-\frac{\sin{2\pi n}}{2\pi n}\right)+\frac{2t}{U}\frac{n\sin(n\pi)}{\pi}\right].        
\label{cc}
 \end{equation}
From the above equation, we readily reproduce Eq. (\ref{graviola}), where 
we noticed that the last term in Eq. (\ref{cc}) is of order $(J/t)$ and 
will be neglected.  
\section{Derivation of Eq.~(\ref{fruta})}
\label{tudo}
Equation (\ref{verdura}) can be written as
 \begin{eqnarray}
 \frac{\partial{\hat{\varepsilon}_k}}{\partial\mu}=2\sum_{k'>0}\frac{f_{k,k'}^{s}\beta
(1-\partial{\hat{\varepsilon}_{k'}}/\partial\mu) e^{\beta(\hat{\varepsilon}_{k'}-\mu)}}{[e^{\beta(\hat{\varepsilon}_{k'}-\mu)}+2]^2}
\nonumber\\
+ 2\sum_{k'<0}\frac{f_{k,k'}^{s}\beta
(1-\partial{\hat{\varepsilon}_{k'}}/\partial\mu) e^{\beta(\hat{\varepsilon}_{k'}-\mu)}}{[e^{\beta(\hat{\varepsilon}_{k'}-\mu)}+2]^2}.
\label{greve}
\end{eqnarray}
 Next we explore the presence of sharp peaks at the Fermi surface:
 \begin{eqnarray}
 \frac{\partial{\hat{\varepsilon}_k}}{\partial\mu}\simeq 2f_{k,k_F}^{s}\sum_{k'>0}\frac{\beta
(1-\partial{\hat{\varepsilon}_{k'}}/\partial\mu) e^{\beta(\hat{\varepsilon}_{k'}-\mu)}}{[e^{\beta(\hat{\varepsilon}_{k'}-\mu)}+2]^2}
\nonumber\\
+ 2f_{k,-k_F}^{s}\sum_{k'<0}\frac{\beta
(1-\partial{\hat{\varepsilon}_{k'}}/\partial\mu) e^{\beta(\hat{\varepsilon}_{k'}-\mu)}}{[e^{\beta(\hat{\varepsilon}_{k'}-\mu)}+2]^2}.
\label{grevista}
\end{eqnarray}
Both integrands are now even, thus after using
$2\sum_{k'>0}(\cdots)=2\sum_{k'<0}(\cdots)=\sum_{k' }(\cdots)$,
 one gets Eq.~(\ref{fruta}) with the help of Eq.~(\ref{verdureiro}).

 \section{Derivation of Eq.~(\ref{liga})}
 \label{cru}
 Deriving $E(\phi)$ with respect to $\phi$, one gets
\begin{eqnarray}
\frac{\partial E(\phi)}{\partial\phi}=\sum_{k,\alpha} 2t^{\ast}\sin(k+\phi)\delta\langle\hat{n}_{k,\alpha}\rangle_{\phi}
\nonumber\\
+\sum_{k,\alpha;k',\alpha'}f_{k,\alpha;k',\alpha'}\delta\langle\hat{n}_{k,\alpha}\rangle_{\phi}\frac{\partial}{\partial\phi}\delta\langle\hat{n}_{k',\alpha'}\rangle_{\phi},
\label{beleza}
\end{eqnarray}
where we have explored the symmetry $f_{k,\alpha;k',\alpha'}=f_{k',\alpha';k,\alpha}$ and
 neglected the exponentially small term
\begin{equation}
W\equiv \sum_{k,\alpha}\left[
\tilde{\varepsilon}_{k,\alpha}(\phi) \frac{\partial}{\partial\phi}\delta\langle\hat{n}_{k,\alpha}\rangle_{\phi}\right].
\label{andes}
\end{equation}
In order to demonstrate this point,
we derive Eq.~(\ref{terreno}) with respect to $\phi$:
 \begin{equation}
 \frac{\partial \hat{\varepsilon}_{k,\alpha}(\phi)}{\partial\phi}=2t^{\ast}\sin(k+\phi)+\sum_{k',\alpha'}f_{k,\alpha; k', \alpha'}
\frac{\partial}{\partial\phi}\delta\langle\hat{n}_{k',\alpha'}\rangle_{\phi}.
\label{caramelo}
\end{equation}
Since $f_{k,\alpha; k', \alpha'} ={\cal O}(t^2/U)$, we can use in (\ref{caramelo}) the approximation
\begin{equation}
\delta\langle\hat{n}_{k',\alpha'}\rangle_{\phi}=\frac{1}{e^{\beta[\varepsilon_{k',\alpha'}(\phi)-\mu]}+2}-
\langle\hat{n}_{k',\alpha'}\rangle_{\substack{\phi=0\\ T=0\\ U=\infty}},
\label{lenda}
\end{equation}
 where $\varepsilon_{k',\alpha'}(\phi)=-2t\cos(k'+\phi)$.
 We note that $\sum_{k',\alpha'}\frac{1}{e^{\beta[\varepsilon_{k',\alpha'}(\phi)-\mu]}+2}=N$ implies
 $\mu=-2t\cos(n\pi)-k_BT\ln{2}+\cdots$, which is $\phi$-independent. Therefore,
 \begin{equation}
 \frac{\partial}{\partial\phi}\delta\langle\hat{n}_{k',\alpha'}\rangle_{\phi}=-\frac{\beta e^{\beta[\varepsilon_{k',\alpha'}(\phi)-\mu]}2t\sin(k'+\phi)}{[e^{\beta[\varepsilon_{k',\alpha'}(\phi)-\mu]}+2]^2}.
 \end{equation}
 After inserting this derivative back into Eq.~(\ref{caramelo}), one has
 \begin{eqnarray}
 \frac{\partial \hat{\varepsilon}_{k,\alpha}(\phi)}{\partial\phi}=2t^{\ast}\sin(k+\phi)
 \nonumber\\
 -\sum_{k',\alpha'}
 f_{k,\alpha; k', \alpha'}
 \frac{\beta e^{\beta[\varepsilon_{k',\alpha'}(\phi)-\mu]}2t\sin(k'+\phi)}
 {[e^{\beta[\varepsilon_{k',\alpha'}(\phi)-\mu]}+2]^2}.
 \label{couve}
 \end{eqnarray}
 We now sum over all values of $\alpha'$ to get an expression that is $\alpha$-independent:
 \begin{eqnarray}
 \frac{\partial \hat{\varepsilon}_{k,\alpha}(\phi)}{\partial\phi}=2t^{\ast}\sin(k+\phi)
 \nonumber\\
 -2\sum_{k' }
 f_{k, k' }^{s}
 \frac{\beta e^{\beta[\varepsilon_{k' }(\phi)-\mu]}2t\sin(k'+\phi)}{[e^{\beta[\varepsilon_{k' }(\phi)-\mu]}+2]^2},
 \label{couves}
 \end{eqnarray}
 with omission of the subscript $\alpha'$ in $\varepsilon_{k',\alpha'}(\phi)$.
 We shall now demonstrate that the above sum is weakly dependent on $\phi$. In the thermodynamic limit, it reads
 \begin{equation}
 I(k)\equiv -\frac{L}{\pi}\int_{-\pi}^{\pi} dk'f_{k,k'}^{s}
 \frac{\beta e^{\beta[\varepsilon_{k'}(\phi)-\mu]}2t\sin(k'+\phi)}{[e^{\beta[\varepsilon_{k'}(\phi)-\mu]}+2]^2},
 \label{bseis}
 \end{equation}
 where it is to be noted that the integrand exhibits sharp peaks at $k'+\phi=\pm k_F$.
 After the transformation $k'+\phi=q$ one obtains
 \begin{equation}
 I(k)= -\frac{L}{\pi}\int {dq} f_{k,q-\phi}^{s}
 \frac{\beta e^{\beta(\varepsilon_{q} -\mu)}2t\sin{q} }{[e^{\beta(\varepsilon_{q} -\mu)}+2]^2},
 \end{equation}
so that the dependence on $\phi$ is removed from the integrand (except for the very small dependence of
 $f_{k,q-\phi}^{s}$) and we can take $\phi=0$ with negligible error [$\mu_0=-2t\cos(n\pi)$]:
\begin{equation}
 I(k)= -\frac{L}{2\pi}\int_{-\pi}^{\pi} {dq} f_{k,q }^{s}
 \frac{\beta e^{\beta(\varepsilon_{q} -\mu_0)}2t\sin{q} }{[e^{\beta(\varepsilon_{q} -\mu_0)}+1]^2},
 \end{equation}
 with limit of integrations restituted to their original values.
 Exploring again the presence of sharp peaks at the Fermi surface (see Appendix \ref{tudo}), we obtain
 \begin{equation}
 I(k)=-\frac{L}{2\pi}( f_{k,k_F }^{s}- f_{k,-k_F }^{s}),
 \label{bnove}
 \end{equation}
 with the use of $|\sin{q}|=\sqrt{1-\cos^2{q}}$. 
 
 We now return to Eq.~(\ref{couve}):
 \begin{equation}
 \frac{\partial \hat{\varepsilon}_{k,\alpha}(\phi)}{\partial\phi}=2t^{\ast}\sin(k+\phi)
 -\frac{L}{2\pi}( f_{k,k_F }^{s}- f_{k,-k_F }^{s}).
 \label{caranguejo}
 \end{equation}
 The derivative of $\delta\langle\hat{n}_{k,\alpha}\rangle_{\phi}$,
Eq.~(\ref{sacola}), with respect to $\phi$ can be now calculated:
 \begin{eqnarray}
 \frac{\partial}{\partial\phi} \delta\langle\hat{n}_{k,\alpha}\rangle_{\phi}=
 -\frac{\beta e^{\beta[\hat{\varepsilon}_{k,\alpha}(\phi)-\mu]}}{[e^{\beta[\hat{\varepsilon}_{k,\alpha}(\phi)-\mu]}+2]^2}
 \nonumber\\ \times
 \left[ 2t^{\ast}\sin(k+\phi)-\frac{L}{2\pi}( f_{k,k_F }^{s}- f_{k,-k_F }^{s})\right].
 \label{galinha}
 \end{eqnarray}

We are now in a position to show that $W$ is exponentially small. 
After inserting Eq.~(\ref{galinha}) into (\ref{andes}), one gets
 \begin{eqnarray}
 W=\sum_{k,\alpha}\left[4(t^{\ast})^2\sin(k+\phi)\cos(k+\phi)
 \right. \nonumber\\ \left.
 -2t^{\ast}\cos(k+\phi)
 \left(\frac{L}{2\pi}\right)(f_{k,k_F}^{s}-f_{k,-k_F}^{s})\right]
 \nonumber\\ \times
 \frac{\beta e^{\beta[\hat{\varepsilon}_k(\phi)-\mu]}}{[e^{\beta[\hat{\varepsilon}_k(\phi)-\mu]}+2]^2}.
 \nonumber\\
 \end{eqnarray}
Once again, we call attention to the fact that the integrand displays sharp peaks at
$k+\phi=\pm k_F$. Thus, after making the transformation $q=k+\phi$,
the resulting integrand becomes odd in $q$. Using the same arguments
that we have applied to go from Eq.~(\ref{bseis}) to (\ref{bnove}),
we thus conclude that $W$ is exponentially small.
We can now return to Eq.~(\ref{beleza}) and derive it one more time with respect to $\phi$:
\begin{eqnarray}
&&\left( \frac{\partial^2 E(\phi)}{\partial\phi^2} \right)_{\phi=0}=
 \sum_{k,\alpha} 2t^{\ast}\sin{k} \left( \frac{\partial}{\partial\phi}\delta\langle\hat{n}_{k,\alpha}\rangle_{\phi}
 \right)_{\phi=0}
 \nonumber\\
 &&+\sum_{k,\alpha} 2t^{\ast}\cos{k} \,\, \delta\langle\hat{n}_{k,\alpha}\rangle_{\phi=0}
 \nonumber\\
&&+\sum_{k,\alpha;k',\alpha'}f_{k,\alpha;k',\alpha'}\left(\frac{\partial}{\partial\phi}\delta\langle\hat{n}_{k,\alpha}\rangle_{\phi}\right)_{\phi=0}
\left(\frac{\partial}{\partial\phi}\delta\langle\hat{n}_{k',\alpha'}\rangle_{\phi}\right)_{\phi=0}
\nonumber\\
&&+\sum_{k,\alpha;k',\alpha'}f_{k,\alpha;k',\alpha'} \,\,\delta\langle\hat{n}_{k,\alpha}\rangle_{\phi=0} \left( \frac{\partial^2}{\partial\phi^2}\delta\langle\hat{n}_{k',\alpha'}\rangle_{\phi}\right)_{\phi=0}.
\label{beldade}
\end{eqnarray}
At low temperatures, we neglect terms containing $\delta\langle\hat{n}_{k,\alpha}\rangle_{\phi=0}$ in (\ref{beldade}), 
and make use of Eq. (\ref{galinha}) and of the procedure that led to Eq. (\ref{bnove}) to 
obtain the final result:
\begin{eqnarray}
\left( \frac{\partial^2 E(\phi)}{\partial\phi^2} \right)_{\phi=0}
=-\frac{2t^{\ast}L}{\pi}\sin(n\pi)
\nonumber\\
+\left(\frac{L}{\pi}\right)^2(f_{k_F,k_F}^{s}-f_{k_F,-k_F}^{s}),
 \end{eqnarray}
which implies Eq. (\ref{liga}).

\end{document}